\documentclass[preprint2]{aastex}

\slugcomment{}

\shorttitle{Suzaku observation of 1H 0419-577}
\shortauthors{Turner et al.}
\def\suzaku{{\em Suzaku}\ }

\begin{document}


\title{Suzaku Observation of a Hard Excess in 1H 0419-577: \\
    Detection of a Compton-Thick Partial-Covering Absorber}

\author{T.J.Turner}
\affil{Department of Physics, University of Maryland Baltimore County, 
   Baltimore, MD 21250 and Astrophysics Science Division,   
NASA/GSFC, Greenbelt, MD 20771, U.S.A}

\author{L.Miller} 
\affil{Dept. of Physics, University of Oxford, 
Denys Wilkinson Building, Keble Road, Oxford OX1 3RH, U.K.}

\author{S.B.Kraemer}
\affil{Institute for Astrophysics and Computational Sciences, Department of Physics, 
The Catholic University of America, 
Washington, DC 20064; and Astrophysics Science Division, NASA Goddard Space Flight Center, 
Greenbelt, MD 20771}

\author{J.N.Reeves}
\affil{Astrophysics Group, School of Physical and Geographical Sciences, Keele 
University, Keele, Staffordshire ST5 5BG, U.K}

\and

\author{K.A.Pounds} 
\affil{Department of Physics and Astronomy, University of Leicester, Leicester LE1 7RH, UK}

\begin{abstract}

We present results from a 200 ks \suzaku observation of 1H 0419-577 taken during 2007 July. 
The source shows a strong excess of counts above 10 keV compared to the extrapolation of  
models based on previous data in the 0.5-10 keV band. 
The 'hard excess' in 1H 0419-577 can be explained by 
the presence of a 
Compton-thick partial-covering absorber that 
covers $\sim $ 70\% of the source. The Compton-thick gas  likely originates  from a radius inside of  
the optical BLR  and may form part of a clumpy disk wind. The fluorescent 
Fe K$\alpha$ luminosity measured by \suzaku is consistent with that expected from an 
equatorial disk wind.

\end{abstract}


\section{Introduction}

1H 0419-577 is a broad-line Seyfert 1 galaxy located at a redshift z=0.104 \citep{grupe96,thomas98,turner99b} with a 
strong ultraviolet flux \citep{marshall95}. The X-ray properties of this AGN have been the subject of 
much discussion, as the source shows a strong steepening to soft energies and marked spectral variability 
\citep{guainazzi98,turner99b,page02}. 
The extreme nature of the spectral variability of this source is especially interesting since  
systematic flattening at low flux levels is a common property of Seyfert galaxies 
 \citep[e.g.][]{papadakis02,pounds04a, pounds04b,vaughan04a, miller07a,miller08a,turner08a}.  
Understanding the origin of the observed phenomenon should lead to important insight as to the 
physical processes that dominate in such sources. 
The  AGN low-state spectra  have been suggested by some \citep[e.g.][]{vaughan04a,miniutti04a} to represent blurred reflection arising within 20${\rm r_g}$ and by others to represent a 
complex absorbed state \citep[e.g.][]{inoue03,pounds03a,miller08a}. In the context of the former model, 
spectral variability has been attributed to 
changes in the degree of light bending as the continuum source location moves; for absorption-based models 
spectral variability is attributed to changes in the ionization, column or covering of 
absorbing gas \citep[e.g.][]{netzer02a,kraemer05a}. 
As for many sources, detailed models for 1H 0419-577  have been suggested whereby the 
spectral form and variability can be explained either by a partial-covering 
absorber \citep{pounds04a,pounds04b} or 
blurred reflection from the inner accretion disk \citep{fabian05}; data prior to \suzaku were 
unable to provide a definitive preference for either of the two scenarios. 

Recent improvements in the quality of data available in the Fe K regime has led to 
the detection in AGN of relatively narrow  absorption lines of high EW  
\citep[$> 50$ eV, e.g.][]{pounds03a,miller07a,dadina05a,braito07a,turner08a}, indicative of an origin in very large columns of ionized gas.
 Narrow absorption lines cannot arise from blurred reflection so  their  
detection motivates a reconsideration of the importance of absorption in shaping the observed 
X-ray properties of AGN. This, and the extended bandpass afforded by the combination of 
XIS and PIN data motivated a \suzaku observation of 1H 0419-577 that is the subject of this paper.

\section{Observations}

\suzaku \citep{mitsuda07} has four X-ray telescopes each  focusing X-rays on to a CCD 
forming part of the X-ray Imaging Spectrometer \citep[XIS][]{koyama07}  
suite. XIS units 0,2,3 are front-illuminated (FI) and  cover $\sim 0.6-10.0$ keV 
with energy resolution FWHM $\sim 150\,$eV at 6 keV. 
Use of XIS2 was discontinued after a charge leak was discovered in 2006 November.
XIS 1 is a back-illuminated CCD and has an enhanced soft-band response (down to 0.2 keV) 
but  lower area at 6 keV than the FI CCDs as well as a larger 
background level at high energies. 
\suzaku also carries a non-imaging, collimated  Hard X-ray Detector  
\citep[HXD][]{takahashi07} whose PIN detector provides useful AGN data typically over 15-70 keV.

The \suzaku  observation was made 2007 July 25 (OBSID 702041010). 
We re-ran the pipeline processing of the raw data to utilize the most recent 
calibration with {\sc hxdpi} and  {\sc hxdgrade} versions 2008-03-03. 
 The  data were then reduced using v6.4.1 of {\sc HEAsoft}. We screened the events 
to exclude data during and within 
500 seconds of  entry/exit from the 
South Atlantic Anomaly (SAA).  Additionally we 
excluded data with an Earth  elevation angle less than 10$^\circ$ and 
 cut-off rigidity $>6$ GeV. The source was observed at the nominal 
center position for the XIS. The FI CCDs were in $3 \times 3$ and $5 \times 5$ 
edit-modes, with normal clocking mode. 
For the CCDs utilized we selected good  
events with grades 0,2,3,4, and 6 and removed hot and flickering  pixels using the 
SISCLEAN script.  The spaced-row charge  injection (SCI) 
was utilized. The exposure time was 179 ks for the FI CCDs.

The XIS products were extracted from 
circular regions of 2.9\arcmin \,   radius while background spectra were extracted from a region 
of the same size offset from the source (and avoiding the chip corners
where calibration source data are registered).  The response and ancillary response
files were then created using {\sc xisrmfgen v2007 May} and {\sc xissimarfgen v2008 Mar}. 
The background was $1\%$ of the total XIS count rate in the 
full XIS band for each CCD.  

1H 0419-577 is too faint to be detected in the HXD GSO instrument, but was 
detectable in the PIN. For the analysis we used the model ``D'' background 
(released 2008 June 17 
\footnote[1]{http://www.astro.isas.jaxa.jp/suzaku/doc/suzakumemo/suzakumemo-2007-01.pdf}).  
As the PIN background rate is strongly variable around the orbit, 
 we first selected source  data to discard events within 500 s of an SAA passage, we also 
rejected events with  day/night elevation angles $> 5^{\rm o}$.
The time filter resulting from the screening was then applied to the background events 
model file to give PIN model-background data for the same time intervals covered by the
 on-source data. The net exposure time was 142 ks. As the 
background events file was generated using ten times the actual
background count rate, an adjustment to the background spectrum was 
applied to account for this factor.  {\sc hxddtcor v2007 May} was run to apply the deadtime 
correction to the source spectrum. To take into account the cosmic X-ray background 
\citep{boldt87,gruber99} {\sc xspec} v 11.3.2ag  was used 
to generate a spectrum 
from a CXB model \citep{gruber99}  normalized to the $34 \times 34 '$ 
{\it Suzaku} PIN field of view, 
and combined
with the PIN instrument background file to create an total background file.  The 
flux of the CXB component is $8 \times 10^{-12} {\rm erg\, cm^{-2}s^{-1}}$ in the 
15-50 keV band. 
The source comprised 15\% of the total counts in the PIN band.  We used the response file 
\verb+ae_hxd_pinxinome3_20080129.rsp+ for spectral fitting. 

Spectral fits utilized data  from the {\sc XIS} instrument, detectors 0 and 3,
in the energy range $0.6-10$\,keV
and also from the {\sc PIN} instrument providing useful data in the 
energy range $15-50$\,keV 
for this weak source. XIS1 was not used owing to the 
higher background level at high energies. 
In the spectral analysis, the {\sc pin} flux was increased by a 
factor 1.16 which is the  appropriate  adjustment for the instrument  
cross-calibration at the epoch of the observation 
 \footnote[2]{ftp://legacy.gsfc.nasa.gov/suzaku/doc/xrt/suzakumemo-2008-06.pdf}. 
Data in the range 1.78-1.9 keV were excluded from the XIS due to 
uncertainties in calibration around the instrumental Si K edge. 
A  2\% systematic error was applied to the PIN background spectrum before spectral fitting. 
XIS  data were binned at the HWHM resolution for each instrument, optimal for detection 
of spectral features while PIN data were binned to be a minimum of 5 $\sigma$ 
above the background level for the spectral fitting.

\section{Spectral Fitting Results}

\subsection{Initial Results}

During our \suzaku observation the source was found to have (background subtracted) count rates 
2.613 $\pm 0.004$ (summed XIS0,3)  and $5.42 \pm0.17 \times 10^{-2}$ (PIN) ct/s corresponding to a 
2-10 keV source flux 
$1.8 \times 10^{-11} {\rm erg\, cm^{-2} s^{-1}}$ and 
0.5-2 keV flux 
$1.4 \times 10^{-11} {\rm erg\, cm^{-2} s^{-1}}$. These fluxes represents a high state as compared to 
the ranges 
$0.9 - 1.6 \times 10^{-11} {\rm erg\, cm^{-2} s^{-1}}$ in the 2-10 keV band  \citep{pounds04a,page02} and 
$0.2 - 1.3 \times 10^{-11} {\rm erg\, cm^{-2} s^{-1}}$ in the 0.5 - 2 keV band 
observed by {\it XMM}. 

The full-band XIS light curve showed only modest variability of $\pm40\%$ around the mean. The PIN-band 
flux did not show any significant variability within the observation. The weakness of the target 
combined with the relatively low level of flux variability meant that it was not possible to extract a useful 
decomposition of the data using Principal Components Analysis \citep{miller07a}. Thus, in this paper we report only upon 
analysis of the mean \suzaku spectrum and application of our model to the older 
 {\it XMM} spectra.

\begin{figure}
\epsscale{.60}
\plotone{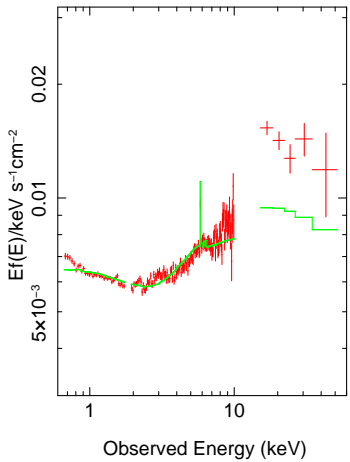}
\caption{\suzaku XIS$0+3$ and PIN data (red) compared to the XMM-based  
absorption model (solid green line). PIN data are rebinned more than  5$\sigma$ above the background 
for the plot only \label{fig:badfit}}
\end{figure}

\citet{pounds04a,pounds04b} developed an absorption-based model for  1H 0419-577 where variations in the 
opacity or covering fraction of the absorber  could explain the marked spectral variability correlated with flux; 
that model had one partial-covering absorber and one full screen of gas intrinsic to the AGN. 
We  constructed a model of that form and a  fit was performed allowing 
 the powerlaw index, normalization, column densities and ionization states all to be free. 
The model also included  a full-covering  
 screen of neutral gas fixed at the Galactic line-of-sight column and the  emission features detected by 
the {\it XMM} 
Reflection Grating Spectrometers  \citep{pounds04b} that fell within our bandpass. 
As shown in Table~1 (row 1), this fit  yielded $\chi^2=631/188$ degrees of freedom ($dof$). The model is unable to 
account for the  excess of counts above 10 keV (Figure ~\ref{fig:badfit}),  where data lie more than 
$\sim 50\%$ above the expected flux.

Before fitting models in more detail we performed some additional analysis.  As the background 
level dominates the 
PIN count rate we  focus on the reliability of the new PIN background model to determine the 
robustness of the PIN result. 
We re-processed the PIN data using the old background model (``A'') and old processing pipeline. This 
combination yields 
a background subtracted flux for 1H 0419-577 that is 8\% lower than that for the new processing with new 
 model D background. Re-fitting the data using the older version of the PIN pipeline 
and background model reduces the PIN excess by 10\% but still leaves a highly significant excess.  
Unfortunately there were no Earth occultation data  to test against the  PIN model 
predicted rates for this date and time of observation. 
To check whether thermal noise in the PIN detector might affect the PIN rate, which can sometimes be a problem 
below  20 keV,  we 
conservatively re-screened the PIN data to  discard events within 5760 s of an SAA passage.
No significant difference was found between the more tightly screened PIN spectrum and  
the nominally processed version so there seems to be no thermal problem with the few PIN channels used below 20 keV. 
Finally, as the 1$\sigma$ uncertainty on the model D background is 1.3\%, we increased the PIN 
background spectrum 
by  3$\sigma$ and found this did not significantly reduce the PIN hard excess. 
The tests performed indicate that the PIN result for this AGN is robust to the precise method of 
 PIN background estimation used. 

The PIN field of view is $34' \times 34'$ and as this is not an imaging instrument we checked 
sky catalogs and  confirmed  that no known hard X-ray sources lie within the PIN 
 field-of-view that could contaminate our spectrum. The {\it ASCA} GIS images are particularly useful 
as these cover  0.7-10.0 keV with a field of view $50'$ diameter. 
{\it ASCA} GIS images 
\footnote[1]{http://tartarus.gsfc.nasa.gov/products/74056000/74056000\_gsfc.html}
taken 1996 July and August show that there are  no other hard X-ray sources detected 
  in the  field of view at that epoch.

1H 0419-577 was detected in the BAT survey \citep{tueller08}. The 
BAT spectrum was downloaded from the public archive 
\footnote[2]{http://heasarc.nasa.gov/docs/swift/results/bs9mon/} and fit with our model to determine 
the flux in a bandpass that can be compared directly to the 2007 PIN data. 
We found  a mean BAT flux  for the period Dec 2005 - Sept 2006 
$F_{15-50}=1.5^{+0.38}_{-0.36} \times 10^{-11} {\rm erg\, cm^{-2}s^{-1}}$,  
lower than the PIN flux over the same band 
$F_{15-50}=2.6\pm0.13 \times 10^{-11} {\rm erg\, cm^{-2}s^{-1}}$.  
Another hard-band measurement was obtained using the {\it BeppoSAX} PDS which gave 
$F_{15-136} = 1.2  \pm0.56 \times 10^{-11} {\rm erg\, cm^{-2}s^{-1}}$ \citep{deluit03} 
compared to the extrapolation of the PIN fit which yields 
$F_{15-136} = 3.5\pm0.2  \times 10^{-11} {\rm erg\, cm^{-2}s^{-1}}$. In conclusion, 
comparison of  measurements  above 10 keV shows some evidence for flux variability and we 
return to this point in the discussion. 

\subsection{Fitting  Absorption-Dominated Models}
\label{absfit}

In the context of absorption models, the hard excess is indicative of the 
presence of a layer of Compton-thick gas  partially covering the continuum source. 
(The continuum  optical depth has  a value of unity for 
$N_H \simeq 1/1.2\sigma_{\mathrm T} \simeq 1.25\times 10^{24} {\mathrm{cm}^{-2}}$ and so 
we refer to a column of gas 
at or above this threshold as Compton-thick). 
To fit the PIN data, we modified the \citet{pounds04a,pounds04b}  
model to allow the second absorber to have a covering fraction less than 1, and 
have a different column density and ionization state to that previously fitted. 

Initially we used a table generated from {\sc xstar} v21ln8 with
gas density assumed to be $n=10^{10}\, {\rm cm^{-3}}$, an illuminating spectrum a powerlaw of index 
$\Gamma=2.3$ and a gas turbulent velocity 300 km s$^{-1}$. The value of assumed turbulent velocity is 
of great importance when one is deriving ionic column densities from fitting narrow absorption lines 
and varying the turbulent velocity over the range 100-500 km s$^{-1}$ can lead to 
changes in the inferred column density by 
an order of magnitude \citep[e.g.][]{young05a}. However, 
the absorber fits here are based upon detection of broad features and overall 
spectral curvature and so are not very sensitive to 
 turbulent velocity assumed in generating the table. 
The fit is shown in Table~1 (errors are only given on the best fit). 
A limitation of {\sc xstar} is that spectral modification from Compton scattering or 
'Comptonization', important for Compton-thick gas, are neglected: in our solution, one zone of gas 
falls at a value where these effects start to become important and so our model for the source is, in 
that sense, an approximation to the  true physical situation. However, as our fit falls so close to the 
Compton-thick boundary, and as many assumptions have had to be made in the fitting, we consider this 
an adequate parameterization for data of this quality.

\begin{figure}
\epsscale{.60}
\plotone{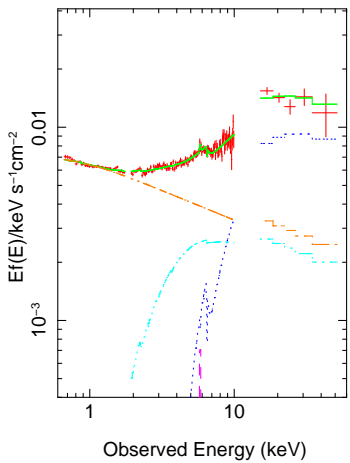}
\caption{\suzaku XIS$0+3$ and PIN data (red) compared to the best partial-covering  model. 
The solid (green) line represents the sum of all model components, the 
 dotted (orange) line represents the uncovered fraction of the continuum, the  dashed (dark blue) 
line is the fraction of 
continuum absorbed by Compton-thick gas,  the  dot-dash (aqua) line is the fraction of continuum under 
$5 \times 10^{22}\, {\rm cm^{-2}}$ and the dashed (purple) line is the Fe K$\alpha$ emission. 
 \label{fig:poundsgoodfit} }
\end{figure}

The powerlaw continuum has three sight-lines, 18\% of the continuum 
is absorbed only by the Galactic 
column, 66\% of the observed continuum has passed through a Compton-thick column 
and 16\% through 
$\sim  5 \times 10^{22}{\rm cm^{-2}}$ (that we dub the 'intermediate layer') and the fit is 
shown in Figure ~\ref{fig:poundsgoodfit}. 
The model was constructed and parameters tabulated 
for a physical  interpretation where  two partial-covering absorbers are co-spatial. 
Obviously, different interpretations of the fit can be made if the gas 
layers are assumed to lie one inside of the other; an obvious possibility is that the 
intermediate layer lies outside of the Compton-thick clouds (constructing the model in 
that way does not significantly change the fitted $\xi$ and $N_H$ values but changes the inferred 
covering fraction for the intermediate layer to 82\%). 
In the discussion we assume the former, co-spatial, geometry. 

The solution we have found for the  Compton-thick zone (Table~1) leads to an expectation of 
weak absorption lines from Fe in the 6.4-6.6 keV regime.  However,    
owing to the low covering fraction of the gas the equivalent width of the 
strongest  absorption line 
would be $< 10$ eV, undetectable in these data. 

The Fe K$\alpha$ line was found to have a peak energy $6.39 \pm 0.06$\, keV, 
$\sigma=0.07 ^{+0.33}_{-0.07}$\, keV,
$n=6.2^{+6.6}_{-2.2} \times 10^{-6}$ photons cm$^{-2}\, s^{-1}$ 
yielding an equivalent width 30 eV against the total observed continuum.  
The combined uncertainties in line width and flux leave this measurement consistent with those 
from {\it XMM} data \citep{pounds04a,pounds04b}.  
The column density returned from the fit is degenerate with $\xi$ and depends on 
the gas density assumed in the {\sc xstar} model run; consequently  the {\sc xstar} table density 
significantly affects the estimate of flux lost to scattering in the Compton-thick gas.   
Fitting the data using a model table with $n=10^{12}\, {\rm cm^{-3}}$  the column density for the 
highest column layer fell to 
9.7$^{+0.64}_{-0.48}\times 10^{23}\, {\rm cm^{-2}}$ with log\,$\xi=0.82^{+0.14}_{-0.21}$. 

\begin{figure}
\epsscale{.60}
\plotone{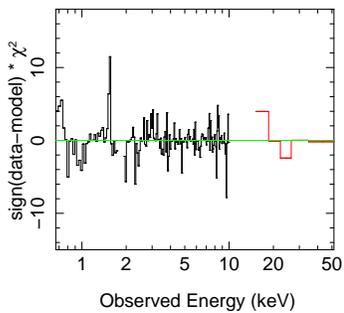}
\caption{The $\chi^2$ residuals to the best X-ray fit shown in Figure 2, from fitting 
 a partial covering model as detailed in  Table~1 
 \label{fig:poundsgoodfitchi} }
\end{figure}

The dominant remaining contribution to $\chi^2$ 
is at 1.55 keV (Figure ~\ref{fig:poundsgoodfitchi}), 
coincident with the Al  detector edge. While the Al edge is much weaker than the Si edge (whose 
peak energy-band is excluded from the fit), the presence of a detector feature at an energy where we see 
an isolated emission 
line of uncertain identification is problematic, and thus we do not attempt to assign an 
astrophysical explanation to the feature. 

We also tried a variation on this model that included a component of (unblurred) reflection linked to 
the Fe K line by a factor 725 in normalization and 
parameterized by the {\sc xspec} model {\sc pexrav}; this linked component pair  physically  
represent reflection from distant neutral material 
illuminated by a  continuum of slope $\Gamma=2.3$ (with a high energy cutoff outside of the 
PIN bandpass) for a disk inclined at 
 $\theta=60^{\rm o}$ to the observer's line-of-sight \citep{george91a}.   However in the fitting 
process we found  
this {\sc pexrav} component to be sufficiently weak that its inclusion 
did not improve the fit or significantly affect the absorption parameters.

The flux observed for  1H 0419-577 during the \suzaku observation corresponds to an observed 
luminosity  
$L_{2-10} = 4.2 \times 10^{44}\, {\rm erg\,s^{-1}}$, 
(assuming $H_0=75\, {\rm km\, s^{-1}\, Mpc^{-1}}$ throughout this paper). 
The absorption-corrected flux and luminosity   are 
$F_{2-10} = 5.3 \times 10^{-11}\, {\rm erg\, cm^{-2} s^{-1}}$, 
$L_{2-10} = 1.3 \times 10^{45}\, {\rm erg\,s^{-1}}$,
(with no significant difference between the  fits using tables of different density gas). 
We also need to account for the scattering losses in the high-column gas zones. 
For the first solution (Table~1) assuming that the filling factor of the gas is low 
(such that no significant flux is scattered back into the line of sight)  then, as the  
column $1.85 \times 10^{24}\, {\rm cm^{-2}}$ 
transmits only 29\% of the incident flux,  a factor 3.4 correction must be applied to the 
fraction of flux observed through that layer to estimate the total true intrinsic luminosity. 
For the alternative model solution, yielding 9.7$\times 10^{23}\, {\rm cm^{-2}}$, the 
scattered flux is half of that incident on the cloud, under the same assumption about filling.  
Correcting the fraction of flux coming through the Compton-thick zone for scattering losses  we thus estimate that 
the intrinsic luminosity for the source must lie in the range 
$L_{2-10} \sim 2 - 3 \times 10^{45}\, {\rm erg\,s^{-1}}$  
and from these values the bolometric luminosity is estimated \citep{elvis94a} to be 
$L_{bol} \sim 7 - 11 \times 10^{46}\, {\rm erg\,s^{-1}}$. 

Finally, there is an intriguing possibility that the underlying continuum is curved, as evidenced by 
examination of the difference spectrum from {\it XMM} data \citep{pounds04b}. As we have not sampled a 
large range of flux in this \suzaku observation we are unable to test this idea further but note that  
a simple extrapolation of the  Comptonized spectral form  found by \citet{pounds04b} 
into the PIN band would increase the implied hard excess.  

\subsection{Extension to {\it XMM} high/low data}
\label{hilo}

  As reported by \citet{page02,pounds04a,pounds04b} the 2000 December 
(OBSID 0112600401) 
and 2002 September (OBSID 0148000201) {\it XMM} observations of 1H 0419-577 sampled the 
source  across a wide range of flux states, ideal for testing the applicability of the \suzaku model to 
other epochs. We reduced the {\it XMM} data following the method and using the criteria of {\citet{pounds04b} 
except that {\sc sas} version 8.0.0 was utilized and only the pn data were extracted.
We fit the spectra fixing the photon index, column densities and ionization states 
of the two absorbing layers from the \suzaku fit using the $n=10^{10}\, {\rm cm^{-3}}$  case as in Table~1. 
Assuming a model construction as before 
we found the high state to yield a solution whereby $18\pm1$\% of the source is uncovered, 
$24\pm2$\% is covered by the intermediate layer of $\sim 5 \times 10^{22}\, {\rm cm^{-2}}$, and
$58^{+21}_{-14}$\%  also lies under the Compton-thick absorber.
Fitting the {\it XMM} low-state data we found that only $3\pm1$\% of the flux was unobscured, while 
$27\pm4$\% was covered by the intermediate column and 
 $70\pm16$\% also covered by the  Compton-thick layer. 
As found for several other sources, and previously suggested for 1H 0419-577 \citep{pounds04b}
 the dominant change in spectral shape observed below 10 keV is consistent with 
changes in the fraction of unobscured continuum, while all data are consistent with 
the same column density and ionization-state for the gas. 
Noting that the energetically-dominant hard band is much less variable than the soft band X-rays 
we examined the absorption- and scattering-corrected fluxes for the low and high-state data and found that  
under the \suzaku  model,  both  
{\it XMM} datasets returned the same implied intrinsic continuum flux of 
$F_{2-10} \sim 5.4 \times 10^{-11}\, {\rm erg\, cm^{-2} s^{-1}}$ 
 showing the continuum source to be consistent with no intrinsic continuum variation 
between {\it XMM} epochs and thus 
attributing observed variations  to  changes  in the absorbers.   
However, the issue of whether the continuum source is intrinsically constant 
remains unresolved: after both absorption and scattering corrections have been applied 
the \suzaku data imply an intrinsic continuum flux 
$F_{2-10} \sim 8 - 15 \times 10^{-11}\, {\rm erg\, cm^{-2} s^{-1}}$, higher than the {\it XMM} values.

\subsection{Fitting Blurred Reflection Models}

The {\it XMM} data for 1H 0419-577 has also previously been  
fit successfully using a composite disk model  \citep{fabian05}
 with relativistic blurring parameterized 
by the {\sc xspec} model {\sc kdblur} \citep{laor91a}. The blurred model is constructed from the sum of a 
powerlaw plus reflection of that continuum from an ionized disk \citep{ross05a}. 
To fit the {\it XMM} spectrum \citet{fabian05} utilized 
 a composite disk, ie one constructed to allow a change in 
 the illumination pattern at a break radius, while the  ionized absorbing gas was parameterized by 
an edge at 0.74 keV. 

We fit the data using  a model with disk inclination, inner and outer radii, break radius, emissivity 
indices and Fe abundance 
fixed at the values found by \citet{fabian05}, 
as these parameters are not expected to vary with 
epoch or source flux.  
The normalization of the composite disk and of the powerlaw, 
$\Gamma$, the ionization parameters and the  absorption edge parameters were all allowed to be free. 
The model was fit to the 
XIS and PIN data and yielded  $\chi^2=1154/199\,  dof$), failing to fit the high PIN flux above 10 keV. 
We then freed the break-radius and both emissivity indices, yielding  $\chi^2=529/189$ $dof$  
with the fit again failing to explain the source flux above 10 keV. 
Further freeing the Fe abundance, disk inclination angle, inner radius  
and adding a narrow Fe emission line component yielded  $\chi^2=337/183$ $dof$. 
Replacement of the absorption edge with a full screen of ionized absorbing gas did not yield a significant improvement. 
As the blurred reflection model does not fit the data we continue with a 
discussion of  the source in the context of the absorption-dominated model. 

\section{Discussion}

\subsection{The location of the absorbing zones}

Data from several wavebands support the existence of partial-covering absorbers in AGN. 
UV spectroscopy reveals multiple zones of gas distinguished by different ionization-states, 
covering fractions and kinematics. Partial-covering by neutral gas and/or ionized absorbing gas 
has long been suggested as one possible way to explain the spectral curvature in AGN 
\citep[e.g][]{reichert85a,holt80a} and the gas was found to be consistent with an origin in 
the optical broad line region  \citep[BLR;][]{holt80a,piro92a}. 
Recent results from X-ray grating spectroscopy have shown that 
multiple layers of gas contribute to  X-ray absorption, including  columns of gas 
with  $N_H \sim 10^{23} - 10^{24}\, {\rm atoms\, cm^{-2}}$ with  
very high ionization-states revealed by their H-like and He-like Fe K absorption lines 
\citep[e.g.][]{young05a, miller07a,turner08a}. In the context of absorption-dominated models, spectral variability 
is often explained by changes in covering or opacity of the gas layers, or a response in  gas 
ionization-state as the continuum varies \citep[e.g.][]{netzer02a,kraemer05a}. Deep minima in the light curves of MCG-6-30-15 
\citep{mckernan98a} and NGC 3516 \citep{turner08a} support a variable-covering absorption picture 
as their dip profiles  suggest an origin as eclipse events. Further support for the 
importance of absorption is seen from the short-timescale absorption line variations detected in 
NGC 1365 \citep{risaliti05a}. 

With the relatively low spectral resolution available in the X-ray regime to date, progress has 
been limited by 
the ambiguity between absorption and blurred reflection signatures: even the brightest and best 
studied AGN such as MCG-6-30-15  can be described using either model.  
\cite[c.f.][]{miniutti03a,miniutti04a,miniutti07a,inoue03,miller08a}. However, the break to a hard excess 
in the \suzaku  PIN data for 1H 0419-577 
 is sufficiently sharp that a preference is found for a partial-covering model: 
the Compton-thick  zone  covers 66\% of the continuum source, 
similar to the covering fractions implied for several other high-column partial-covering absorbers 
 \citep[e.g.][]{reeves02a,reeves03a,miller07a,turner07a}. 

A common argument against partial-covering models is based on probability, i.e. that 
to have a covering fraction such as  $\sim 70\%$ with significant observed variations in covering 
(here of order 10\% between flux states)  implies that there must be obscuration from a few clouds 
subtending a similar transverse size on the sky as the continuum source. Obviously, 
if the clouds exist as far out 
as the BLR then the probability of observing an orbiting cloud cross the line-of-sight is low.   
However, the distances estimated for absorbing clouds depend on the cloud density assumed.  
If one assumes a high density for the absorber (e.g. $n \gtrsim 10^{12} {\rm cm^{-2}}$) then 
the material is indicative of an origin  close to the active nucleus and the 
probability issue is alleviated. 
Further to this, a likely scenario is that the material  exists in the 
form of a disk wind rather than 
discrete clouds \citep[e.g.][]{sim08a,schurch08a}, 
in that case the argument about probability  is not relevant  and the 
observation of two partial absorbers reveals the clumpy nature of the wind. 
In any case, some of the unabsorbed flux may be scattered emission, the 3\% residual unabsorbed flux 
observed in our interpretation of the low-state {\it XMM} data  (Section~\ref{hilo}) seems likely 
to be such (allowing for this, of course, increases the inferred cloud covering fraction).  

Here, the distance of the absorbers can be estimated from the definition of $\xi$ used 
in {\sc xstar}, $\xi=\frac{L}{r^2n}$, where 
$L$ is the ionizing luminosity integrated from
1-1000 Ry, $n$ is the proton density and $r$ the distance of
the material from the central black hole \citep{tarter69a}. 
We took the column, ionization and implied intrinsic luminosity from fits assuming 
gas densities $n=10^{10}\,{\rm cm^{-3}}$ and $n=10^{12}\,{\rm cm^{-3}}$ 
to obtain a range of radial estimates for the Compton-thick region 
$r_{CT} = 5 \times 10^{16}\, {\rm cm} - 2 \times 10^{17}\, {\rm cm}$. 
Both the scaling relations of \citet{kaspi00a} and \citet{bentz06} yield 
an estimate $r_{BLR} \sim 2 \times 10^{17}$cm for the radius of the BLR using the 
5100 \AA\, flux from  \citet{guainazzi98}. Based on this the Compton-thick 
partial-covering absorber appears to lie within the BLR.   

1H 0419-577 provides one of  few clear examples to date  of a  Compton-thick  
partial-covering absorber. PDS~456 exhibits a similar 
 hard excess (at the 3 $\sigma$ level) in \suzaku  data to that seen here 
(Reeves et al. 2009, submitted). {\it RXTE} data for PDS~456 support the 
\suzaku result, showing a very deep Fe K  edge from  gas of column 
density $\sim 5 \times 10^{23}\, {\rm cm^{-2}}$, $\log\xi \sim 2.5$ flowing outward at 
 50,000\,km\,s$^{-1}$ \citep{reeves02a,reeves03a}. 
The question of gas velocity is interesting, high velocity outflows in the range 
0.1--0.2~c \citep{reeves08a} have been reported in 
X-ray data for a number of high luminosity AGN 
\citep[e.g][]{reeves02a,pounds03a,dadina04a,dadina05a,chartas03a} and we 
suggest that 1H 0419-577 has a high velocity outflow whose absorption features 
have yet to be isolated. 

Interestingly, model fits to  {\it BeppoSAX} data showed 1H 0419-577 to have a flat 
component in the hard 
X-ray regime \citep[modeled in that case as a $\Gamma \sim 1.5$ powerlaw, ][]{guainazzi98}, although 
the hard-band flux is different to that observed by the PIN. If
 the continuum source is intrinsically constant, then the apparent variability in flux 
over 15-136 keV  
can be achieved only if there exists  a component of  material  with 
$N_H > 10^{25} {\rm cm^{-2}}$ along the line-of-sight: the implication of such an extension to high 
column is not  surprising given the establishment of clumps of material in the 
regime $N_H \sim \times 10^{24}\, {\rm cm^{-2}}$ found here. 

\subsection{An origin for the absorption in a disk wind associated with the BLR}

Unfortunately the \suzaku  data for 1H 0419-577 yield little insight into the 
kinematics of the X-ray absorbers. However, 
examination of the UV spectra has proved interesting in this regard. 
\citet{dunn07a} show the O {\sc vi} and Ly$\beta$ absorption lines to have kinematic components tracing outflow 
at 200, 100 and 20 km s$^{-1}$, with evidence that the lowest velocity component of gas does not fully cover the source. 
An {\it IUE} spectrum  
of 1H 0419-577 (Figure ~\ref{fig:civ}) shows 
a C {\sc iv} emission line with an apparent blue wing. This phenomenon is 
commonly observed in the high-ionization lines in quasar spectra 
and generally  attributed to a lack of flux on the red side of the emission line \citep{richards02a,murray98a}. 
In models where  the high-ionization UV emission 
lines are thought to arise from clouds comprising a disk wind, this 
profile may be indicative of a disk viewed close to edge-on, such that 
the observer is looking down the flow of the wind  \citep[e.g.][]{murray98a, richards02a}. In the case of 
1H 0419-577 the C {\sc iv} profile has a  velocity width $\sim 5000\, {\rm km\, s^{-1}}$ 
(Figure ~\ref{fig:civ}) which is consistent 
with the widths of the O {\sc vii} and O {\sc viii} lines  of 7000$\pm 3000 {\rm km\, s^{-1}}$  
detected by \citet{pounds04b}.  
The width of the Fe K$\alpha$ emission, $\sim 7000^{+9000}_{-7000} {\rm km\, s^{-1}}$,   detected here is consistent with the UV and 
soft X-ray results, indicating all these signatures arise in  kinematically-associated 
flows. \citet{baskin05} suggest 
that high $\frac{L}{L_{Edd}}$ is a necessary (but not sufficient) condition for generating a blueshifted 
C {\sc iv} profile. 
The mass of the central black hole in this source has been estimated at 
$1.3 \times 10^8$ M$_{\odot}$ (Pounds et al 2004b);  
taking 
$L_{bol} \sim 7 - 11 \times 10^{46}\, {\rm erg\,s^{-1}}$ we  estimate that 
1H 0419-577 is operating at $L \sim 3 - 6  \times L_{Edd}$.  Of particular interest is the 
prediction of Compton-thick, clumpy, continuum-driven disk winds for Eddington-limited AGN \citep{king03a}; 
this suggests that similar 'hard excesses' will be found  for other super-Eddington 
AGN.

\begin{figure}
\epsscale{.80}
\plotone{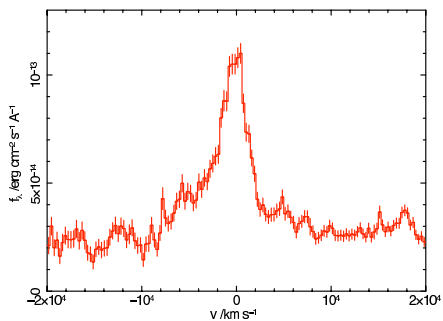}
\caption{IUE spectrum of 1H 0419-577 showing the asymmetric C {\sc iv} profile
 \label{fig:civ}}
\end{figure}

\subsection{The bolometric luminosity of 1H 0419--577}

In the optical regime 1H 0419-577 has observed  line strength 
 \citep{guainazzi98} 
$F_{H \alpha} = 1.6 \times 10^{-12}\, {\rm erg\, cm^{-2}\, s^{-1}}$ (broad component) 
with corresponding luminosity  
$L_{H \alpha} = 3.8 \times 10^{43}\, {\rm erg\,s^{-1}}$.
The correlation established between H$\alpha$ and L$_{\rm x}$ \citep{ho01,ward88} 
allows us to estimate the line expected for gas exposed to a given luminosity. 
Taking the estimate of  intrinsic (scattering corrected)  2-10 keV luminosity from our partial-covering model, 
$L_{2-10} = 2 - 3  \times 10^{45}\, {\rm erg\,s^{-1}}$ and the known line-luminosity correlation  we 
would predict  
$L_{H \alpha} \sim 0.2-2 \times  10^{44}\, {\rm erg\,s^{-1}}$ and so 
the observed  H$\alpha$ line luminosity is consistent with the expected value. The 
observed O {\sc iii} $\lambda 5007$ line luminosity is 
$L_{O III} = 1.4 \times 10^{43}\, {\rm erg\,s^{-1}}$ \citep{grupe99}. 
\citet{melendez08a}  find a relationship between O {\sc iii} and X-ray luminosity that 
yields an expected O {\sc iii} luminosity 
$L_{O III} = 0.8-3 \times 10^{43}\, {\rm erg\,s^{-1}}$, consistent with that observed. These good 
agreements between expected and observed line fluxes suggest our derivation of intrinsic continuum luminosity is 
reasonable, supporting our application of the absorption and scattering corrections utilized in 
regard to the  source flux.

\subsection{Fluorescent line emission from the absorbing zones}

The  absorbing gas is, naturally, expected to 
produce fluorescent line emission.  
The line intensity may depend on a number of factors, 
the most obvious being the fraction of the source continuum that is intercepted by 
the absorber.  This line could be significant from the Compton-thick layer of gas. 
We should also distinguish between two cases, one where the fluorescent
emission experiences resonant scattering, and one where it does not.

Considering  Fe\,K$\alpha$ emission: at the value of $\xi$ inferred from the fits in 
section\,\ref{absfit}, $0.5 \la \log\xi \la 2$, for photon index $\Gamma=2.3$, we do not
expect there to be a significant ion  population more highly ionized than Fe {\sc xvii}  
\citep{kallman04a} and, in this case, this line would not experience substantial resonant scattering.
We can then infer approximately the expected line luminosity by multiplying the emission 
from the spherically-symmetric {\sc xstar} absorber model by some filling factor.

The predicted model flux in the Fe K$\alpha$ line is $L_{Fe} \sim 3^{+3}_{-1} \times 10^{43}\,{\rm erg\, cm^{-2}\, s^{-1}}$ 
and the observed luminosity is $\sim 5^{+5}_{-1.5}\%$ of that predicted for a full shell of gas. Assuming that the disk 
extends beyond the radius where the gas exists (applicable to the disk wind scenario) then the 
optically-thick disk will hide half of the emission from view (thus the true  global covering for the gas  
 is $\sim 10^{+10}_{-3}$\% and we denote this $G_f=0.1$). 
In the case where the absorber is a smooth equatorial wind seen edge-on we thus estimate a disk
opening angle $\theta \simeq \sin^{-1}G_f \simeq 12^{\circ}\, ^{+12}_{-4}$ \citep[c.f.][who found $G_f \sim 0.3$ 
for PG1211$+143$]{pounds08a}.  

How reasonable is such a wind opening angle?  If we had surveyed a sample of AGN which
had been selected in a way independent of orientation, we would expect to be
viewing about 10\,percent of them through the equatorial wind.
Such an unbiased sample has not been investigated from this perspective, but
this probability of occurrence  implies that the inferred source
orientation of 1H 0419--0577 is not particularly surprising.

However, given the uncertainty in the ionization of the Compton-thick zone, 
and given the simplicity of the model adopted,
we cannot exclude the possibility that this zone has $\log\xi \ga 2.5$, in which
case charged states more highly ionized than Fe {\sc xviii}  should be populated and
Fe\,K$\alpha$ is expected to suffer resonant scattering.  Resonant scattering
will significantly deplete the emitted line photons, by the Auger effect (for Li- to F-like states) 
and by photoelectric absorption, on repeated scatterings \citep[as discussed by e.g.][]{turner09a}. 

In this case 
the geometry of the absorber and its orientation with
respect to the observer play an even more key role than in the non-resonant case
\citep[e.g][]{ferland92a}:
most line photons will be scattered and escape through the upper and lower surfaces 
of the disk. If we are viewing the disk edge-on, most of those photons will be lost
from the line-of-sight.  Hence in the resonant scattering case, extremely large
line suppression factors are possible, in which case the estimated filling factor deduced
assuming no resonant scattering is very much a lower limit to the true filling factor.

\section{Conclusions}

A marked 'hard excess' of counts  is 
detected in the \suzaku  PIN data for 1H 0419-577 relative to the predicted flux based on 
fits below 10 keV. The \suzaku data can be fit using an absorption-dominated model 
showing that  Compton-thick partial-covering X-ray absorbers  exist in the type 1 AGN, 
1H 0419-577 and probably other AGN. 
 The data are consistent with a clumpy disk wind that 
provides the X-ray absorption from gas residing at radii inside of the BLR. The observed 
Fe K$\alpha$ line emission is consistent with an origin in an equatorial disk wind 
and the widths of measured  soft X-ray and UV emission lines  support this picture.  
We find that blurred reflection models cannot satisfactorily fit the 
source and so the 1H 0419-577  data provide a rare distinction between  
two observationally-similar classes of model.

\acknowledgments

TJT acknowledges NASA grant NNX08AL50G. LM acknowledges STFC grant number PP/E001114/1. We thank the anonymous referee for comments that helped improve 
this manuscript. 
We are also grateful to the \suzaku operations team  for performing  this observation and providing 
software and calibration for the data analysis. Some of the data presented in this paper were 
obtained from the Multi-mission Archive at the Space Telescope Science Institute (MAST). STScI 
is operated by the Association of Universities for Research in Astronomy, Inc., under NASA 
contract NAS5-26555. Support for MAST for non-HST data is provided by the NASA Office of 
Space Science via grant NAG5-7584 and by other grants and contracts.
This research has also made use of data obtained from the High Energy Astrophysics Science 
Archive Research Center (HEASARC), provided by NASA's Goddard Space Flight Center.



\bibliographystyle{apj}      
\bibliography{xrayreview_feb11}   

\begin{deluxetable}{cccccccc}
\tabletypesize{\scriptsize}
\rotate
\tablecaption{Partial Covering Model}
\tablewidth{0pt}
\tablehead{
\colhead{$\Gamma $} & 
\colhead{$N_H1$ \tablenotemark{1}} & 
\colhead{log $\xi 1$} & 
\colhead{C1 \tablenotemark{2} } & 
\colhead{$N_{H}2$ \tablenotemark{1} } &
\colhead{log $\xi 2$} & 
\colhead{C2 \tablenotemark{2} } & 
\colhead{$\chi^2/dof$} 
}
\startdata
2.19 & 0.66 & -3.0 & 48\%   &  0.03  & 4.1 & 100\%  & 631/188 \\
2.31$\pm0.04$ & 0.54$^{+0.11}_{-0.21}$  & -0.13$^{+0.80}_{-1.09}$ & 16$\pm3$\% &  
   18.5$^{+1.9}_{-3.7}$ & 1.90$^{+0.16}_{-0.49}$ & 66$^{+12}_{-3}$\%  & 260/187 \\
\enddata
\tablecomments{A column of neutral gas covered all components, fixed at the Galactic 
value 2.0 $\times 10^{20}{\rm cm^{-2}}$. Errors are calculated 
at 90\% confidence but only shown for the best fit}
\tablenotetext{1}{Column density in units of 10$^{23} {\rm atom\, cm^{-2}}$}
\tablenotetext{2}{Percentage Covering}

\end{deluxetable}

\end{document}